\documentclass[fleqn,usenatbib]{mnras}

\usepackage[english]{babel}
\usepackage{graphics,graphicx}
\usepackage[usenames]{color}
\usepackage[authoryear]{natbib}

\usepackage[T1]{fontenc}
\usepackage{ae,aecompl}
\usepackage{txfonts}
\usepackage{verbatim}

\newcommand{\be}{\begin{equation}}
\newcommand{\ee}{\end{equation}}
\newcommand{\bdm}{\begin{displaymath}}
\newcommand{\edm}{\end{displaymath}}
\newcommand{\bea}{\begin{eqnarray}}
\newcommand{\eea}{\end{eqnarray}}
\newcommand{\ba}{\begin{array}}
\newcommand{\ea}{\end{array}}

\newcommand{\pa}[1]{\left(#1\right)}
\newcommand{\paq}[1]{\left[#1\right]}
\newcommand{\ds}{\displaystyle}

\title{Gravitational waves from neutron star excitations in binary inspirals}

\author[A. Parisi, R. Sturani]{Alessandro Parisi$^{1}$ and Riccardo Sturani$^{2}$\thanks{Contact e-mail: riccardo@iip.ufrn.br}
\\
$^{1}$ICTP-South American Institute for Fundamental Research, 
Instituto de F\'isica Te\'orica (UNESP),  01140-070  S\~ao Paulo, Brazil\\
$^{2}$International Institute of Physics (IIP), Universidade Federal do Rio Grande do Norte (UFRN) CP 1613, 59078-970  Natal-RN, Brazil}

\begin{document}

\maketitle

\begin{abstract}
In the context of binary inspiral of mixed neutron star - black hole systems,
we investigate the excitation of the neutron star oscillation modes by the
orbital motion.
We study generic eccentric orbits and show that tidal interaction can excite the
$f$-mode oscillations of the star by computing the amount
of energy and angular momentum deposited into the star by the orbital motion
tidal forces via closed form analytic expressions.
We study the $f$-mode oscillations of cold neutron stars using recent
microscopic nuclear equations of state, and we compute their imprint into the
emitted gravitational waves.
\end{abstract}


\section{Introduction}
After the historical detections of gravitational waves by binary black holes
\cite{TheLIGOScientific:2016pea},
it is expected that mixed binaries composed of a neutron star (NS) and
a black hole (BH) may be the next, qualitatively different type of source to be
detected in the gravitational wave (GW) channel.
At first approximation mixed NS-BH can be treated in
General Relativity (GR) on equal footing as binary BH systems, however the
presence
of matter in the GW source may lead to new detectable astrophysical
effects in the GW signal that are not expected to appear in the binary BH case
like e.g.
NS tidal deformations leaving an imprint in the GW signal
\cite{1992ApJ...400..175B,Flanagan:2007ix} and 
breaking of the NS giving origin to a gamma ray burst or more
general electromagnetic counterpart \cite{1976ApJ...210..549L}, to name
only the most studied effects. \\
Beside their direct phenomenological relevance, these effects carry information
on the highly uncertain equation of state of the NS, thus making GW detection an 
invaluable probe of the internal structure of NSs.
In this work we focus on a specific effect in GW signals: NS can be tidally
deformed by the orbital motion in generic elliptic orbits, hence setting
oscillations of the NS normal modes. The orbit being elliptical can induce
resonant oscillations at a frequency much higher than the frequency
scale set by the inverse of the orbital period, since in general NS
oscillations are much higher than orbital frequency of inspiral binary systems.\\
Quantifying this phenomenon in light of the exciting prospect of a future
GW detection has been the subject of extensive investigations in literature in
a number of different contexts.
The theoretical setup for studied such tidally induced NS oscillations has been
provided in \cite{Thorne:1969rba,PressTeu}. In \cite{1975MNRAS.172P..15F} it
was originally proposed that tidal encounters between a NS and a main-sequence
star might lead to the formation of X-ray binaries in globular clusters.
In \cite{Shibata:1993qc} the effects of the tidal resonances for a
\emph{circular} orbital motion has been studied,
with the result if the companion of a NS is a
BH of mass $\geq 6M_\odot$, the $g$-mode resonance is unimportant, while the
$f$-mode resonance may affect the orbital evolution just before the merging.
\cite{Rathore:2004gs} considered the energy absorbed by tidal excitations in
\emph{eccentric} orbit (but not their imprint in the GW-form).
\cite{ReiGol} compute the effect on the emitted GW phase of resonant mode
excitation by the \emph{circular} inspiral motion.
Rotating NS we considered by \cite{Ho:1998hq} (including  g-modes and r-modes)
when the spin axis is aligned or anti-aligned with the orbital angular momentum 
axis. \cite{1983A&A...121...97C} solved for the tidal deformation dynamics
of a NS in an external field of a massive object
and recently \cite{Chirenti:2016xys} presented a framework for the discussion of binary 
NS and mixed NS-BH ones oscillation mode excitation and detection via the
GWs observed by future GW detector as Einstein Telescope or Cosmic Explorer.
Numerical results on the GW emission of tidally excited NS oscillations
in the last stages of a coalescence have been given in \cite{Gold:2011df},
and in \cite{Steinhoff:2016rfi}
the imprint of resonant tidal on the gravitational
waveform has been computed within the effective one body description of the
two body orbital motion.\\
In the present paper we consider non-rotating NS with four different equations
of states \cite{Akmal:1998cf,Douchin:2001sv,Walecka:1974qa,Bethe:1974gy} with
the goal of translating resonant excitations of various
$f$-modes for NSs inspiraling binary NS-BH systems that move in an
elliptical orbit into quantitative prediction for the emitted GW-form.\\
Numerical simulations show that most of the energy released in gravitational
waves is indeed transferred into $f$-modes, which are characterized by a 
wave-function free of nodes along the radial direction.
We do not study the possibilities of exciting the g-modes because these modes
are related to the presence of density discontinuities in the outer envelopes
of NSs, see \cite{1987MNRAS.227..265F} and \cite{1993ApJ...417..273S}, 
density discontinuities in the inner core as a consequence of phase
transitions at high density, as studied in \cite{2002PhRvD..65b4010S},
and/or thermal gradients as for a proto-NS, see e.g.~\cite{Ferrari:2002ut}.
In this paper we do not consider the possibility of having discontinuities of
the density, moreover we focus on barotropic equations of state
where the pressure depends only on the energy density, implying
that all g-modes degenerate to zero frequency, hence we focus on the
excitations of $f$-modes.
Our study is based on the following simplifying assumptions:\\
(i) we neglect BH rotation, thus we treat the BH as a point
particle with mass $M_{\rm{BH}}$;
(ii) the  hydrodynamic stability  of NS is computed using the Oppenheimer-Volkoff equations,
but we use Newtonian equations to calculate the oscillation modes, see
Appendix \ref{sec:appA};
(iii) the NS does not rotate and we neglect viscous effects.\\
By implementing the formalism presented in \cite{Thorne:1969rba,PressTeu} we find
generic analytic expressions for the energy and angular momentum
deposited into NS oscillations during the elliptic orbital motion, allowing to
compute the mass quadrupole which is sourcing GW emission, and eventually
comparing it with the orbital quadrupole.\\
The outline of this paper is as follows: in Sec.~\ref{sec:orbital_motion} we
present the setup of the physical system under consideration, 
and we provide new analytic expressions for the dynamics of tidally induced
NS oscillations, which are the main result of this paper.
In Sec.~\ref{sec:gw} we analyze quantitatively their GW emission.
Finally, conclusions for future detectability of NS oscillations in the GW
channel are drawn in Sec.~\ref{sec:conclusions}.  
We set the speed of light $c=1$ throughout this paper.

\section{Coupling of neutron star oscillation modes to orbital motion}
\label{sec:orbital_motion}
\begin{figure}
\includegraphics[width=\linewidth]{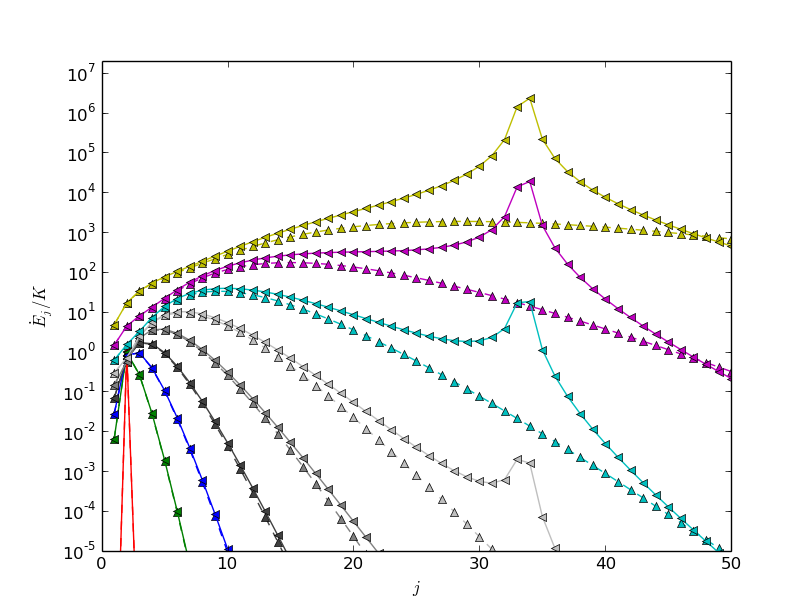}
\caption{Distribution of the energy per unit of mass absorbed by the
fundamental NS oscillation mode divided by the quantity $K$
(defined in eq.~(\ref{eq:Kdef})) as a function of the harmonic of the
fundamental mode $j$ in eccentric orbits for 9 equally spaced values of
eccentricity
(from $e_0=0$ in red, through $e_i=i/10$ until $e_8=0.8$ in yellow).
For each value of eccentricity two curves are reported, for $x=0.01$ and
$x=0.07$ ($x$ defined in eq.~(\ref{eq:xdef})).
For the largest value of $x$ the resonant absorption peaks are visible for
$e=0.5, 0.6, 0.7, 0.8$, as
$\bar j \omega_0= \bar j x^{3/2}/(G_NM)\simeq 
18.1\, \rm{kHz}\, (\bar j/34) (6.965 M_\odot/M) (x/0.07)^{3/2}$
where for this plot $M_{BH}=5M_\odot$ and we used the equation of state A (APR) of
\protect\cite{Akmal:1998cf} and central density $\rho_0=1.5\times 10^{15}$
gr/cm$^3$, see tab.~\ref{table:1}. In this case the $n=0$ $f-$mode
has frequency $\nu_f^{\ell=2}=2.888$ kHz
(we have verified that for $x<0.07$ the NS is safe from tidal braking, whose
condition requires
$\mathcal{D}\lesssim 0.3 R_{NS} (M_{NS}/M_{BH})^{1/3} x^{1/2} (M/M_{BH})^{1/2}$, see
\protect\cite{Vallisneri:1999nq}).
Plots for the other equations of states described in App.~\ref{sec:appC} are
shown in fig.~\ref{fig:edotj_eos} and are qualitatively similar.}
\label{fig:edotj}
\end{figure}

In this section we study the tidal excitation of NS oscillation
modes in non-rotating stars in an elliptical orbit. Our analysis will be general,
but the astrophysical case we have in mind is that of a binary NS-BH system. 
The idea to compute the energy deposited in stellar oscillations by the tidal
gravitational field is first described by \cite{1977ApJ...216..914T}
and \cite{PressTeu}.\\
In this paper we use Newtonian linearized equations to calculate the
oscillation modes. The use of Newtonian equations is consistent with our
Newtonian description of tidal interactions. For the $f$-mode, general
relativistic effects are expected to modify our results of oscillation
frequencies by not more than $GM_*/(R_* c^2)\sim 20$ per cent, see \cite{1994MNRAS.270..611L},
where $M_*$ and $R_*$ are the mass and radius of the NS. We also neglect the spin
$\Omega_s$ of the NS.
When $\Omega_s \neq 0$, the normal modes of the star get more complicated,
especially when $\Omega_s$ becomes comparable to the mode frequencies
\cite{Gaertig:2008uz}.
For $\Omega_s \equiv 0$ the eigenmodes can be adequately approximated by those
of a non-rotating spherical star, the basic equations that governing the
oscillations of stars are discussed in more detail in Appendix \ref{sec:appA}.\\
The NS oscillations are excited by tidal forces while the NS is bound in a
binary system with black hole in an eccentric orbit
whose evolution is driven by gravitational radiation.
The distance $\mathcal{D}$ between two objects in an elliptic orbit can be
parametrized by, see e.g. eq.~(4.54) of
\cite{Maggiore},
\be
\mathcal{D}=\frac{a(1-e^2)}{1+e\cos\psi}
\ee
being $a$ the semi-major axis and $e$ the eccentricity (with $\psi=0$
corresponding to the periastron), and the \emph{true anomaly} $\psi$
is related to the \emph{eccentric anomaly} $u$ and time $t$ via,
see e.g. eqs.~(4.57,58) of \cite{Maggiore},
\be
\label{eq:anomalies}
\ba{c}
\beta \equiv u-e\sin u =\omega_0 t\,,\\
\ds \cos\psi=\frac{\cos u -e}{1-e\cos u}\,,
\ea
\ee
being $T$ the orbital period, $\omega_0\equiv 2\pi/T$ with the
following relationships holding among orbital parameters
\be
\ds\dot\psi=\frac{\paq{G_NMa(1-e^2)}^{1/2}}{\mathcal{D}^2}\,,
\ee
(where $M$ is the total mass of the binary system and $G_N$ the Newton constant)
and the standard definition of the relativistic orbital parameter
\be
\label{eq:xdef}
x\equiv\pa{G_N M\omega_0}^{2/3}=\frac{G_NM}a\,,
\ee
the last equality holding only at Newtonian level.\\
In order to study quantitatively the effect of the
gravitational force inducing oscillations into the NS
and following the procedure outlined in \cite{PressTeu}, it is useful
to expand the Newtonian potential in spherical harmonics, see e.g. eq.~(3.70) of
\cite{Jack}, centered at the star as per
\be
\label{eq:potExp}
\frac 1{|\mathcal{D}-r|}=\sum_{\ell=0}^{\infty}\sum_{m=-\ell}^\ell\frac {4\pi}{2\ell+1}\frac{r^\ell}{\mathcal{D}^{\ell+1}}
Y_{\ell m}^*(\theta,\phi)Y_{\ell m}(\pi/2,\psi)\,,
\ee
being $r,\theta,\phi$ coordinates of the mass elements of the NS,
$\ell,|m|\leq \ell$ are the spherical harmonic indices and the orbital motion
is assumed to be planar (no spin-induced precession).
Using eq.~(\ref{eq:potExp}) for elliptic orbit, it will be useful to expand
$e^{im\psi}/\mathcal{D}^{\ell +1}$ for generic $\ell$ into a Fourier series of the
type
\be
\label{eq:expRphi}
\frac{e^{im\psi}}{\mathcal{D}^{\ell +1}}=\frac 1{a^{\ell+1}}
\sum_{j=0}^{\infty}  \{ c^{(\ell,m)}_j(e)  \cos(j\beta) +
i\; s^{(\ell,m)}_j(e)  \sin(j\beta) \}\,.
\ee
The detailed calculation of the Fourier coefficients
$c^{(\ell,m)}_j(e),s^{(\ell,m)}_j(e)$ and their analytic expressions are presented
in Appendix \ref{sec:appB}.\\
In order to perform an analytic quantitative analysis we borrow here the
framework of \cite{Rathore:2004gs}, where NS oscillations are modeled as a
series of damped harmonic oscillator displacements $x_n(t)$ driven by external
force, that we can take purely monocromatic:
\be
\label{eq:dampedho}
\ddot x_n(t) +2\frac{\dot x_n(t)}{\tau_n}+\omega_n^2 x_n(t)=C_j\cos(\omega_j t)+
S_j\sin(\omega_j t)\,,
\ee
where $\omega_n$ is the stellar mode frequency, $\tau_n$ its damping time\footnote{As a possible mechanism for the damping of non-radial NS oscillations we take the gravitational emission, we do not consider neutrino losses, radiative heat leakage, and magnetic damping.},
$\omega_j\equiv j\omega_0$ is the $j$-th harmonic of the main orbital angular frequency $\omega_0$, and
$C_j,S_j$ the exciting force amplitude.\footnote{Note that the time scale of
$\omega_j$ variation is set by the GW radiation and via the Einstein quadrupole formula
$\frac{\dot\omega_0}{\omega_0}\simeq \frac{96\eta}{5}\pa{G_NM}^{5/3}\omega_0^{8/3}
\rightarrow\frac{\dot\omega_j}{\omega_j}\ll\omega_j$ (as $G_NM\omega<1$, with
$\eta\equiv M_*M_{BH}/M^2$), hence we neglect the time
variation of the frequency of ``forcing'' term in eq.~(\ref{eq:dampedho}).}
Eq.~(\ref{eq:dampedho}) admits the exact analytic solution
\be
\ba{rl}
\left[(\omega_j^2-\omega_n^2)^2\right.&\!\!\!\!\!\left.+4\omega_j^2/\tau_n^2\right]x_n(t)=\\
=&\pa{\omega_n^2-\omega_j^2}\pa{C_j\cos(\omega_j t)+S_j\sin(\omega_j t)}\\
+&2\omega_j/\tau_n\pa{C_j\sin(\omega_j t)-S_j\cos(\omega_j t)}\,,
\ea
\ee
the solution $x_n^{(h)}$ to the homogeneous equation being
\be
x_n^{(h)}\propto e^{-t/\tau_n}\cos\paq{\pa{\omega_n^2-1/\tau_n^2}^{1/2}t+\phi_0}\,,
\ee
leading to an average absorbed energy per unit of mass $\mathcal{E}$ per unit of
time
\be
\label{eq:edot}
\dot \mathcal{E}=\frac{\pa{C_j^2+S_j^2}\omega_j^2/\tau_n}{(\omega_j^2-\omega_n^2)^2+4\omega_j^2/\tau^2_n}\,.
\ee
The NS oscillation vectors $\vec\zeta(t,\vec r)$ satisfy an equation of the
type see \cite{1992MNRAS.258..715K}
\be
\label{eq:osc}
\pa{\rho\frac{d^2}{dt^2}+\mathcal{L}}\vec\zeta(t,\vec r)=
-\rho\vec\nabla U(\vec r)\,,
\ee
where $\mathcal{L}$ is an operator characterizing the internal restoring force
of the star.
In order to apply this toy model of a damped harmonic oscillator to the tidally
excited NS oscillation, we decompose the oscillation field 
$\vec\zeta(t,\vec r)$ into normal modes with factorized time and space
dependence:
\be
\label{eq:zeta}
\vec\zeta(t,\vec{r})=\sum_{n,\ell,m}q_{n\ell m}(t)\vec\xi_{n\ell m}(\vec r)\,,
\ee
where we have added the spherical harmonics $\ell,m$ labels and the spatial
mode eigenfunctions $\xi_{n\ell m}$ satisfy
\be
\pa{\mathcal{L}-\rho\,\omega_n^2}\vec\xi_{n\ell m}=0\,,
\ee
allowing the identification of $\omega_n$ with the stellar frequency of
the eigenmode.
The differential equations the oscillation modes fields $\xi$ satisfy are summarized in Appendix \ref{sec:appA},
which are solved for 4 different equations of
state and 4 values of the central density of the NS, with the resulting
mass, radius, frequency and damping times (the last two depending on $\ell$)
are reported in Appendix \ref{sec:appC} for $2\leq \ell\leq 4$.\\
It is also useful to expand the eigenmodes into a radial ($r$) and a poloidal
($h$) component 
\be
\label{xidef}
\vec\xi_{n\ell m}(\vec{r})=
\left(\xi_{n\ell}^{(r)}(r)\hat e_r+ r\,\xi_{n\ell}^{(h)}(r)\vec\nabla\right)
Y_{\ell m}(\theta,\phi)\,,
\ee
and impose the normalization condition\footnote{Note that with the
normalization chosen $\xi_{n\ell}^{(r,h)}$ have dimension of length, $q_{n\ell m}$ is
dimension-less. However the normalization can be arbitrarily chosen without
affecting physical results, our choice has the advantage of making following
formulae simpler.}
\be
\ba{rcl}
&&\ds\int d^3x\,\rho(r)\,\vec\xi_{n\ell m}^{\ast}\cdot\vec\xi_{n'\ell' m'}\\
&=&\ds
\int dr\,r^2\rho(r)\,\left(\xi_{n\ell}^{(r)}\xi^{(r)}_{n'\ell'}+
\ell(\ell+1)\xi_{n\ell}^{(h)}\xi_{n'\ell'}^{(h)}\right)\delta_{\ell,\ell'}\delta_{m,m'}\\
&=&
\rho_0R_*^5\delta_{n,n'}\delta_{\ell,\ell'}\delta_{m,m'}\,,
\ea
\ee
where $\rho(r),\rho_0,R_*$ are respectively the density, central density
and radius of the NS, and we used
\be
\ba{rcl}
\int d\Omega\, Y_{\ell m}(\theta,\phi) Y^*_{\ell'm'}(\theta,\phi)&=&
\delta_{\ell,\ell'}\delta_{m,m'}\\
\int d\Omega\, r^2\,\vec\nabla Y_{\ell m}(\theta,\phi)\cdot
\vec \nabla Y^*_{\ell'm'}(\theta,\phi)&=&\ell(\ell+1)\delta_{\ell,\ell'}\delta_{m,m'}\,,\\
\int d\Omega\,\vec r\cdot\vec\nabla Y_{\ell m}(\theta,\phi)\,
\vec r\cdot\vec\nabla Y^*_{\ell'm'}(\theta,\phi)&=&\delta_{\ell,\ell'}\delta_{m,m'}\,,
\ea
\ee
and the integral of products of spherical harmonics with unequal number of
derivatives vanish for any $\ell,m$, $\ell',m'$.\\
By multiplying both members of eq.(\ref{eq:osc}) by $\rho(r)\xi_{n\ell m}^*(\vec r)$,
substituting the expansion in eq.~(\ref{eq:potExp}),
and integrating over the NS volume the mode $q_{n\ell m}(t)$ is singled out and it
satisfies an equation of the type (\ref{eq:dampedho}):
\be
\label{eq:qnlm}
\ba{l}
\ds\ddot q_{n\ell m}(t)+\frac 2{\tau_{n\ell}}\dot q_{n\ell m}(t)+\omega_n^2q_{n\ell m}(t)=
\frac{G_NM_{BH}}{a^3}\pa{\frac{R_*}a}^{\ell-2}Q_{n\ell}W_{\ell m}\\
\ds\times\sum_j\pa{c_j^{(\ell +1,m)}(e)\cos(j\beta)+is^{(\ell +1,m)}_j(e)\sin(j\beta)}
\ea
\ee
where
\be
\ba{rcl}
W_{\ell m}&\equiv&\ds\frac{4\pi}{2\ell+1}Y_{\ell m}(\pi/2,0)\,,
\\
Q_{n\ell}&\equiv&\ds\frac 1{\rho_0R_*^{\ell+3}}
\int_0^{R_*} dr\, r^2\rho(r) \ell r^{\ell-1}\pa{\xi^{(r)}_{n\ell}+(\ell+1)\xi^{(h)}_{n\ell}}\,,
\ea
\ee
$M_{BH}$ is the black hole mass.
Note that the r.h.s of eq.(\ref{eq:qnlm}) is complex, but given the symmetries
of the $c,s$ coefficients: $W_{\ell m}=(-1)^\ell W_{\ell -m}$
(and $W_{\ell m}=0$ if $\ell,m$ have different parity),
$c_j^{\ell,m}=c_j^{\ell,-m}$, $s_j^{\ell,m}=-s_j^{\ell,-m}$
the sum of $\sum_m q_{n\ell m}\times Y_{\ell m}$ returns a real quantity.
The modes $q_{n\ell m}$ thus satisfy an equation of the type (\ref{eq:dampedho})
with the coefficients $C_j,S_j$ replaced by
\be
\pa{C_j,S_j}\to \frac{G_NM_{BH}}{a^3}\pa{\frac{R_*}a}^{\ell-2}Q_{n\ell}W_{\ell m}
\pa{c_j^{(\ell,m)}(e),s_j^{(\ell,m)}(e)}\,.
\ee
These expressions will be needed in sec.~{\ref{sec:gw}} to compute the
time varying quadrupole associated to these oscillations, source of
GWs.\\
The rate of energy (per unit of mass, per unit NS radius) absorbed by each
oscillation modes can be read from eq.~(\ref{eq:edot}) by inserting the above
values of $C_j,S_j$, summing over $n, j>0, \ell\geq 2$ and $|m|\leq \ell$, the
rate of absorbed energy via tidal mechanism $\dot E_*$ being
\be
\ba{rl}
\ds\dot E_*=\sum_j\dot E_j=\rho_0R_*\pa{\frac{R_*}a}^4\pa{\frac{G_NM_{BH}}a}^2
&\ds\sum_{j,n,\ell,m}\pa{{c_j^{(\ell,m)}}^2+{s_j^{(\ell,m)}}^2}
\pa{\frac{R_*}{a}}^{2\ell-4}\\
\times&\ds Q_{n\ell}^2W_{\ell m}^2
\frac{\omega_j^2/\tau_{n\ell}}{(\omega_j^2-\omega_{n\ell}^2)^2+4\omega^2_j/\tau_{n\ell}^2}\,.
\ea
\ee
The  contribution from individual $j$ modes to the rate of energy
absorption is plotted in fig.~\ref{fig:edotj} after being divided by the factor
\be
\label{eq:Kdef}
\ba{rcl}
K&\equiv&\ds
\rho_0 R_*\pa{\frac{R_*}a}^4\pa{\frac{G_NM_{BH}}a}^2
\frac{\pa{G_NM\omega_0}^2}{\omega_{0\,2}}\\
&\simeq&\ds 1.5\cdot 10^{-14}\frac{M_\odot}{sec}\pa{\frac x{0.01}}^9
\pa{\frac{\rho_0}{10^{15}\rm{gr/cm^3}}}\pa{\frac{R_*}{10\rm{Km}}}^5
\pa{\frac{M_{BH}}{4 M_{\odot}}}^2\pa{\frac{M}{6 M_\odot}}^{-6}\,,
\ea
\ee
where $\omega_{0\,2}=\omega_{n\ell}$ for $n=0$, $\ell=2$.
Factorizing the absorbed energy rate by the quantity
$K$ has the virtue of making $\dot E_j/K$ dimension-less
and independent on the relativistic parameter $x$ (as long as the
orbital frequency does not hit a resonance with $\omega_j\equiv j\omega_0$)
and mildly dependent on $\rho_0$, $a$.

\begin{figure}
\includegraphics[width=.9\linewidth]{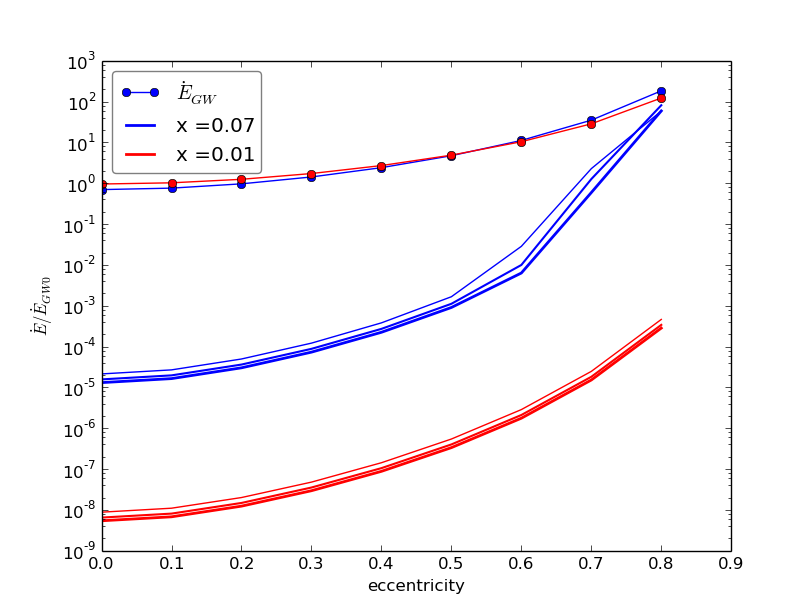}
\caption{Rate of energy absorbed $\dot E_*$ as a function of eccentricity, with
$M_{BH}=5M_\odot$, NS with equation of state A (APR) of
\protect\cite{Akmal:1998cf} for different values 
of the central density $\rho_0=(1.5,1.2,0.99)\times 10^{15}$gr/cm$^3$,
lines of increasing thickness shows results for increasing $\rho_0$.
For comparison we also plot the GW luminosity for the two values of $x$,
all functions are divided by the Newtonian GW luminosity at zero eccentricity
$\dot E_{GW0}$ given by eq.~(\ref{eq:Egw0}).
Plots for the other equations of states described in App.~\ref{sec:appC} are
shown in fig.~\ref{fig:edote_eos} and are qualitatively similar.}
\label{fig:edote}
\end{figure}

In fig.~\ref{fig:edote} we report the absorbed energy rate $\dot E_*$
normalized by
\be
\label{eq:Egw0}
\dot E_{GW0}\equiv \frac{32}{5G_N}\eta^2 x^5
\ee
(being $\eta\equiv M_*M_{BH}/(M_*+M_{BH})^2$ the reduced mass of the orbital system),
which is the expression of the leading order in $x$ of the GW emission rate at zero
eccentricity from a binary inspiral, making visually easier the comparison
between GW radiated energy $\dot E_{GW}$ and $\dot E_*$.
For $\dot E_{GW}$ we use the 3PN formula taken from \cite{Arun:2007sg}, see also
sec.~10.3 of \cite{Blanchet:2013haa}.

\begin{figure}
\includegraphics[width=.9\linewidth]{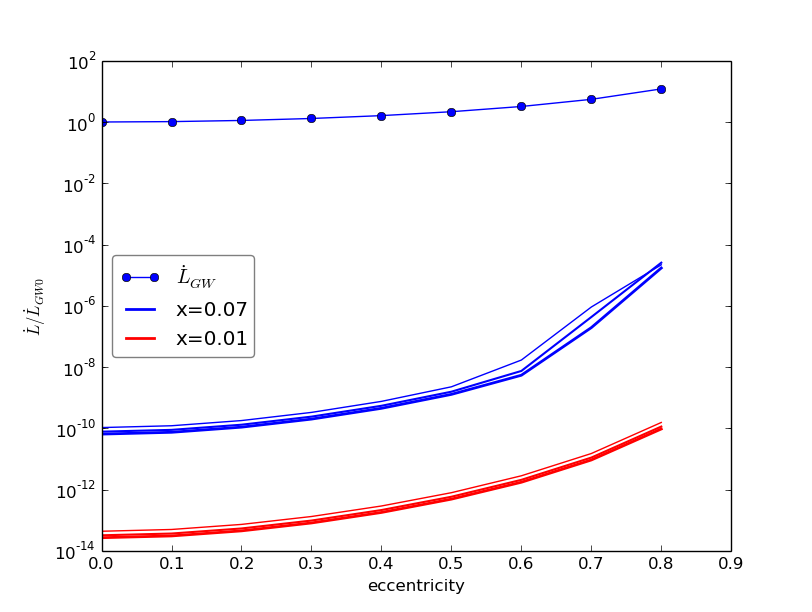}
\caption{Rate of angular momentum absorbed as a function of eccentricity, same
parameters as in fig.~\ref{fig:edote}. Here $\dot L_{GW}$ is the Newtonian
angular momentum loss in GWs for small eccentricities
$\dot L_{GW}=\frac{32}5\eta^2 M\frac{x^{7/2}}{(1-e^2)^2}\pa{1+\frac 78 e^2}$
and $\dot L_{GW0}=\dot L_{GW}|_{e=0}$.}
\label{fig:ldote}
\end{figure}
The absorbed angular momentum can be computed in a similar way, following
\cite{1994MNRAS.270..611L}, where it is noted that the variation of angular
momentum
\be
\ba{rcl}
\ds\dot L_{*}&=&\ds
-\int d^3x(\rho_0+\delta\rho)(\hat z\cdot \vec{r}\times\vec{\nabla}U)
\ea
\ee
we can derive in our setup
\be
\ba{rcl}
\dot L_*
&=&\ds\sum_{n\ell}q_{n\ell m}(t)\int d^3x\vec{\nabla}\cdot(\rho_0\vec{\xi}_{n\ell m})
\frac{\partial U}{\partial\psi}\\
&=&\ds \sum_{n\ell m}q_{n\ell m}(t)\int d^3x\vec{\nabla}\cdot(\rho_0\vec{\xi}_{n\ell m})\frac{G_NM_{BH}}{a}\pa{\frac r{a}}^\ell W_{\ell m}i m\\
&&\ds\times
Y_{\ell m}^*(\theta,\phi)
\sum_j \pa{c_j^{(\ell,m)}(e)\cos(j\beta)+i s_j^{(\ell,m)}(e)\sin(j\beta)}\,,
\ea
\ee
where in the last passage we have inserted the expansion of
eqs.~(\ref{eq:potExp},\ref{eq:expRphi}) and derived by parts inside the integral.
In this form the angular momentum absorption rate by NS oscillations can be
rewritten as:
\be
\ba{rrl}
\dot L_{*}=&\ds\rho_0 R_*\pa{\frac{G_NM_{BH}}a}^2&\ds 2\sum_{j,n,\ell,m>0}
 m\, c_j^{(\ell,m)}s_j^{(\ell,m)}\pa{\frac{R_*}a}^{2\ell}\\
&\times &\ds Q_{n\ell}^2W_{\ell m}^2
\frac{\omega_j/\tau_{n\ell}}{\pa{\omega_j^2-\omega^2_{n\ell}}^2+
4\omega_j^2/\tau_{n\ell}^2}\,.
\ea
\ee
In fig.~\ref{fig:ldote} the absorbed angular momentum rate $\dot L_*$ normalized
by the leading order expression in $x$ of $\dot L_{GW0}\equiv 32/5 M\eta^2x^{7/2}$
is reported for various values of the relativistic parameter $x$ and 
the eccentricity $e$.
The values of $\dot L$ are negligible with respect to $\dot L_{GW}$ and
given the typical moment of inertia of a NS ($\sim 10^{45}$ gr cm$^2$, see book of \cite{Haensel:2007yy}), the induced rotation on the NS is also negligibly small.

\section{Gravitational Wave emission}
\label{sec:gw}
\begin{figure}
\includegraphics[width=\linewidth]{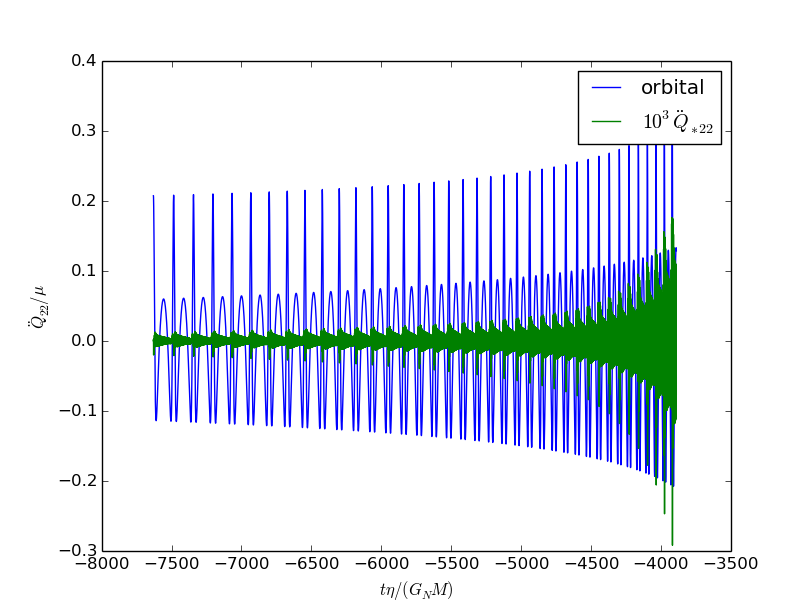}
\caption{Second derivative of the quadrupole $\ddot Q_{22}$ divided by the reduced mass
  $\mu\equiv\eta M$: contribution from
  orbital dynamics compared with (magnified) contribution from the NS oscillation
  $Q_{*22}$ for an inspiral with initial conditions $x_i=0.04$, $e_i=0.4$,
  $M_{BH}=5 M_\odot$ and parameter for the NS given by equation of state B (SLy4) \protect\cite{Douchin:2001sv}
with $\rho_0=2\times 10^{15}$ gr/cm$^3$.}
\label{fig:ddQ}
\end{figure}

We have seen in the previous section that the energy absorbed in by the NS
is very small compared to the orbital energy at moderate eccentricity values
($e\lesssim 0.6$), hence such absorption will not
alter in any significant way the chirping signal.
However the energy absorbed will set oscillations in the neutron star that
gives rise to a time varying quadrupole, which will in turn generate GWs with
a significantly different pattern that the GWs associated to the decaying
orbital motion.\\
The general expression for the GW in the TT gauge is given by,
see e.g. eq.~(3.275) of \cite{Maggiore}, 
\be
h_{ij}^{TT}(t,r)=\frac 1rG_N\sum_{\ell= 2}^{+\infty}\sum_{m=-\ell}^{\ell}\paq{u_{\ell m}\pa{T^{E2}_{\ell m}}_{ij}+
v_{\ell m}\pa{T^{B2}_{\ell m}}_{ij}}
\ee
where $u_{\ell m}$ ($v_{\ell m}$) is linearly related to the $\ell$-th time
derivative of the mass (momentum) multipole moments.
The leading-order contribution to radiation reaction comes from the mass
quadrupole term, for which it is (see e.g. sec.~3 of \cite{Maggiore})
\be
u_{2m}=\frac{16}{15}\pi\sqrt{3}\ddot Q^{ij} {\mathcal{Y}^{2m}_{ij}}^*\,,
\ee
being $\mathcal{Y}^{\ell m}_{i_1\ldots i_\ell}$ the tensor spherical harmonics and
$Q^{ij}\equiv \int d^3x \rho x^i x^j$ is the standard quadrupole mass moment in
Cartesian coordinates.
It will be convenient to express the leading order GW amplitude in terms of the
spherical components $Q_m$ of the quadrupole, related to their
Cartesian counterpart via
\be
\ba{rcl}
\ds Q_{2m}&\equiv&\ds\frac{8\pi}{15}Q_{ij}\pa{\mathcal{Y}^{2m}_{ij}}^*\,,\\
\ds Q_{ij}&=&\ds\sum_{|m|\leq 2} Q_{2m}\mathcal{Y}^{2m}_{ij}\,,
\ea
\ee
leading to (explicit expressions of $\ell=2$ tensor spherical harmonics are
reported in app.~\ref{sec:appD})
\be
\label{eq:uQ}
u_{2m}=2\sqrt 3\ddot Q_m\,.
\ee
We now have all the ingredients to relate the leading GW source $u_{2m}$ to
the NS tidal oscillations via
\be
Q_{*2m}=\frac{8\pi}{15}\int \rho\; r^2 Y_{2\,m}^*d^3x\,,
\ee
that in terms of the displacement vector introduced in
eqs.~(\ref{eq:osc},\ref{eq:zeta}) can be expressed as,
see \cite{Ushomirsky:2000ax}, by
\be
\ba{rcl}
\ds\frac{15}{8\pi} Q_{*2m}&=&\ds\int (\rho_0+\delta\rho) r^2Y_{2 m}^*d^3x\\
&=&\ds -\sum_{n}\int \vec\nabla\cdot\pa{\rho_0\vec\zeta_{n 2m}} r^2 Y^*_{2m}d^3x\\
&\simeq&\ds q_{0 2m}(t)\pa{2\int_0^{R_*} \rho_0\left\{\xi_{0\, 2}^{(r)} +3\xi_{0\, 2}^{(h)}\right\}
r^3 dr-\left.\rho_0\xi_{0\, 2}^{(r)}r^{4}\right|_0^{R_*}}\,,
\ea
\ee
where an integration by parts has been performed in the last step, the
explicit expression of $\vec\zeta_{n2 m}(t,\vec x)$ has been inserted
and only the $n=0$ contribution has been considered.
since we analyzed only the $f$-mode.
Observing that the boundary term is numerically smaller than 
the integral term, substituting the solution of eq.~(\ref{eq:qnlm}) and
considering only the resonant contribution for $\omega_{\bar j}\simeq \omega_{02}$
the NS average quadrupole value can be written as
\be
\ba{rcl}
\ds\langle Q^2_{*22}\rangle^{1/2}&\simeq&\ds
\frac{2\sqrt 2\pi}{15}\rho_0R_*^5Q^2_{02}W_{22}\frac{G_NM_{BH}}{a^3}\frac{\tau_{02}}{\omega_{\bar j}}
\paq{\pa{c^{(2,2)}_{\bar j}}^2+\pa{s^{(2,2)}_{\bar j}}^2}^{1/2}\\
&\simeq&
\ds \frac{4\sqrt 2\pi}{15}\frac{\pa{\rho_0 R_*^5\tau_{02}}^{1/2}}{\omega_{02}}
Q_{02}\sqrt{\dot E_{\bar j}^{(\ell=2)}}
\\
&\simeq&\ds 10^{-2}M_{\odot}\rm{km^2}\pa{\frac{\rho_0}{10^{15}\rm{gr/cm^3}}}^{1/2}\pa{\frac{R_*}{10\rm{km}}}^{5/2}\\
&&\ds\times
\pa{\frac{\dot E_j^{(l=2)}}{10^{-8}M_\odot/\rm{sec}}}^{1/2}
\pa{\frac{\tau_{02}}{0.1\rm{sec}}}^{1/2}\pa{\frac{\omega_{02}}{18\rm{kHz}}}^{-1}
\,.
\ea
\ee
The quantity directly related to GW emission, $u_{2m}^{(NS)}$, follows
straightforwardly via eq.~(\ref{eq:uQ}).
In fig.~\ref{fig:ddQ} we report the contribution to the second time
derivative of the quadrupole (divided by the reduced mass of the binary system)
and as a comparison the (magnified) second derivative of the quadrupole
associated to $n=0$, $\ell =2$ NS oscillations during an ordinary binary 
inspiral in which the orbit shrinks due to GW back reactions.

\begin{figure}
\includegraphics[width=\linewidth]{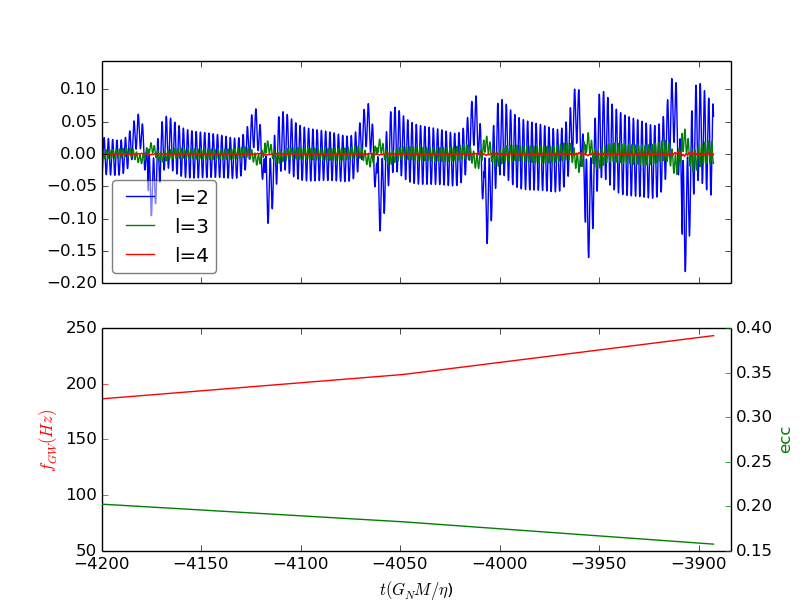}
\caption{Given the same parameters of fig.~\ref{fig:ddQ}, here are displayed the
  $f$-mode displacements $q_{0\ell\ell}$ for $\ell=2,3,4$ (magnified by a factor $10^3$).
Also shown are the main gravitational wave frequency $f_{GW}\equiv \omega_0/\pi$
and the eccentricity along the inspiral dynamics considered.}
\label{fig:ddq*}
\end{figure}
For comparison, we also report in fig.~\ref{fig:ddq*} the time evolution of the
displacement $q_{0\ell \ell}$ $\ell=2,3,4$ along the inspiral phase.

\section{Conclusions}
\label{sec:conclusions}
In this paper we have developed and presented a framework able to perform analytic and quantitative
study of the excitations of a neutron star in an inspiralling binary system of arbitrary eccentricity.
We have computed the energy and the angular momentum deposited into stellar mode oscillations by the tidal
field via closed form analytic formulae.
The amount of energy absorbed by the neutron star in a given mode depends on the overlap
of the tidal force field with the displacement field of the mode, hence it requires solving the
equilibrium equations of a neutron star, done here in the Newtonian approximation.
We focused our analysis on the fundamental $f$-mode of a non-relativistic star,
finding the rate of energy absorbed and angular momentum
as a function of eccentricity and of the period of the inspiral orbital,
when $f$-mode can be in resonance with higher harmonics of the
main orbital frequency.

As a future development of this work, 
we intend to extend our analysis to the General Relativistic equilibrium
equations of a rotating neutron star, with the inclusion of $r$-mode and
$g$-modes, and considering a not barotropic equation of state: such modes have
lower frequency values than the $f$-mode, and can therefore be excited at
resonance in an elliptical orbit earlier in the inspiral phase.
The phenomenological impact of the computations presented here relies on the
signature that neutron star oscillations will imprint onto the gravitational
signals of an inspiral binary system.
Despite being sub-dominant with respect to the gravitational wave
sourced by the orbital motion, the detailed features of the star oscillation
bears invaluable information on its equation of state and density, allowing to
make a bridge to the nuclear physics ruling its equilibrium.
Since it is expected in the near future that third generation gravitational wave detector
could observe signals from binary systems
involving neutron star at signal-to-noise ratio of order $10^2$ or more, see
e.g.~\cite{Punturo:2010zz}, 
and that such detection will involve the observation of hundred of thousand
gravitational wave cycles during the inspiral of a binary system for a time
stretch of order of several days,
the quantitative prediction of the modification of the inspiral
signal, even at very low level, will have an impact on the physics outcome of the detection.

\section*{Acknowledgments}
The authors wish to thank C.~Chirenti for useful discussions.
The work of AP has been supported by the FAPESP grant 2016/00096-6, RS has been supported by FAPESP grant 2012/14132-3.

\bibliographystyle{mnras}

\begin{onecolumn}

\appendix

\section{ Four first-order linear differential equations of non-radial oscillations  }
\label{sec:appA}
The normal modes of a spherical star can be labeled by  spherical    harmonic indices $\ell$ and $m$, and by a ``radial quantum number"  $n$.
In spherical coordinates the Lagrangian displacement $\xi$ of a fluid element is given by 
\begin{equation}\label{NM1}
    \xi_{n\ell m}= \left[\xi_{n\ell}^{(r)}(r), \; \xi_{n\ell}^{(h)}(r) \frac{\partial}{\partial\theta},  \; \frac{\xi_{n\ell}^{(h)}(r)}{\sin\theta} \frac{\partial}{\partial\phi}\right] Y_{\ell m}(\theta,\phi) e^{i\sigma t}
\end{equation}
where  $Y_{\ell m}$ denotes a spherical harmonic; and $\sigma$ denotes the  pulsation angular frequency. 
The oscillation is assumed to be adiabatic,  we ignore the thermal evolution of the NS, for simplicity we use the Newtonian description in the \cite{1971AcA....21..289D} formulation, 
in this case the equations reduce to a system of four first-order differential equations with four dimensionless variables, given by:
\begin{equation}\label{NM2}
 \;\;\; y_1=\;\frac{\xi_{n \ell}^{(r)}}{r}, \;\;\;\;\;\;\;\;\;\;\;\;\; \;\;\;\;y_2=\;\frac{1}{g r}\left(\frac{p'}{\rho}+\Phi'\right)=\;\frac{ \sigma^2}{g} \xi_{n \ell}^{(h)} ,
\end{equation}

\begin{equation}\label{NM3}
\!\!\!\!\!\!\!\!\!\!\!\!\!\!\!\!\!\!\!\!\!\!\!\!\!\!\!\!\!\!\!\!\!\!\!\!   y_3=\;\frac{\Phi'}{gr}, \;\;\;\;\;\;\;\;\;\;\;\;\;\;\;\;\;\;\;  y_4=\;\frac{1}{g }\;\frac{d\Phi'}{dr},
\end{equation}

\noindent Here, the meanings of the symbols are as follows: $p'$ and $\Phi'$ are the radial part of the Eulerian perturbation to the pressure $p$ and the gravitational potential $\Phi$, respectively; $r$ is the distance from the center of the star, $\rho$ is the density, and $g\equiv Gm(r)/r^2$ is the local acceleration due to gravity.
The system of differential equations that governs the linear adiabatic oscillations of stars is then given by:

\begin{eqnarray}
  r \frac{dy_1}{dr} &=&  \left(V_g-1-\ell\right)y_1+\left[\frac{\ell(\ell+1)}{c_1\; \omega^2}-V_g\right]y_2+V_g\;y_3  \\
  r \frac{dy_2}{dr} &=& (c_1\; \omega^2 -A^{\ast})y_1+(3-U+A^{\ast}-\ell)y_2-A^{\ast}y_3 \\
  r \frac{dy_3}{dr} &=& (3-U-\ell)y_3+y_4  \\
  r \frac{dy_4}{dr} &= &  A^{\ast} U y_1+UV_gy_2 + \left[\ell(\ell+1)-UV_g\right]y_3-(U+\ell-2)y_4 \label{NM4}
\end{eqnarray}

Where
\begin{equation}\label{NM5a}
    V_g=-\frac{1}{\Gamma_1}\frac{d \ln p}{d \ln r}=\frac{ g\; r}{c_s^2}, \;\;\;\;\;\;\;\;\;\;\;\;\; A^{\ast}=\frac{1}{\Gamma_1}\frac{d \ln p}{d \ln r}-\frac{d \ln \rho}{d \ln r}, \;\;\;\;\;\;\;\;\;\;\;\;\;\;\;\;U\equiv \frac{d \ln m(r)}{d \ln r}=\frac{4\pi\rho r^3}{m(r)},
\end{equation}

\begin{equation}\label{NM5b}
\!\!\!\!\!\!\!\!\!\!\!\!\!\!\!\!  c_1 \equiv \frac{r^3}{R_{\ast}^3}\frac{M_{\ast}}{m(r)}, \;\;\;\;\;\;\;\;\;\;\;\;\;\;\;\;\;\;\;\; \Gamma_1=\left(\frac{\partial \ln p}{\partial \ln \rho}\right)_{S}, \;\;\;\;\;\;\;\;\;\;\;\;\;\;\;\;\;\;\;\;\;\;\;\;\;\;\; \omega^2=\frac{R_{\ast}^3}{G_NM_{\ast}}\sigma^2\,.
\end{equation}
Here $\Gamma_1$ is the first adiabatic exponent, $c_s$ is the sound speed, $m(r)$ is the concentric mass, $M_{\ast}$ and $R_{\ast}$ are the total mass and radius of the star, respectively, and $G_N$ is the gravitational constant.
There are four boundary conditions, the inner boundary conditions at $r=0$ are: 
$$
\left\{
\begin{array}{rl}
 c_1\;\omega^2 y_1-\ell y_2=0   \\
\ell y_3-y_4=0 
\end{array}
\right.\,,
$$
the outer boundary conditions at $r=R_{\ast}$ are:
$$
 \left\{
\begin{array}{rl}
 y_1-y_2+y_3=0  \\
(\ell +1)y_3+y_4=0
\end{array}
\right.\,.
$$
\noindent The two central boundary conditions require that the two divergences
involved, $ \nabla \cdot \xi_{n \ell}^{(r)}$, $ \nabla \cdot  \Phi'$, remain
finite.
At the surface we require $\delta P/P$ to be finite and $\Phi'$, the
gravitational force per unit mass, to be continuous across the perturbed
surfaces.
The above equations and boundary conditions constitute an eigenvalue problem
for the eigenvalue $\sigma$.\\
The expression for the damping time due to emission of gravitational waves in the Newtonian case  see \cite{Thorne:1969rba};  \cite{1982MNRAS.200P..43B} ) is given by:
\be
\tau_{n\ell} \equiv \frac{(\ell-1)[(2\ell+1)!!]^2}{\ell(\ell+1)(\ell+2)} \left(\frac{\sigma}{2\pi G}\right)\left(\frac{c}{\sigma}\right)^{2\ell+1}
\frac{\int_0^{R_{\ast}}  dr \rho\; r^{2}[\xi_{n\ell}^{(r)}(r)^2+\ell(\ell+1)\xi_{n\ell}^{(h)}(r)^2] }{\left\{ \int_0^{R_{\ast}}  dr \rho\; r^{\ell+1}[\xi_{n\ell}^{(r)}(r)+(\ell+1)\xi_{n\ell}^{(h)}(r)]\right\}^2}
\ee
where  $n!!= 1\cdot\cdot (n-4)(n-2)n$.

\section{Expansion of  the Fourier coefficients }
\label{sec:appB}
Expanding in eq.~(\ref{eq:expRphi}) we have
\be
\ba{ rcl}
c^{(\ell,m)}_j(e)&=&\ds\frac{c_j}{\pi(1-e^2)^{\ell+1}}
\int_{-\pi}^\pi \cos(m\psi)\pa{1+e\cos\psi}^{\ell+1}\cos(j\beta)d\beta\,,
\ea
\ee
for $\ell\geq 0, |m|\leq l$, where we used that $\psi$ is an odd function of
time,
hence $\cos\psi$ ($\sin\psi$)is an even (odd) function of time, $c_j=1$ for
$j\neq 0$, and $c_0=1/2$.\\
In order to expand $\cos(m\psi(t))$ into sums of terms of the type
$\cos(n\beta)$ it is useful to express it in terms of powers of
$\cos(\psi)$ via \cite{AS}
\be
\cos(m \theta)=T_m(\cos(\theta))\,,
\ee
where $T_m$ is the Chebyshev polynomial of order $r$ and it has the form
\be
T_m(x)=\sum_{k=0}^{[m/2]}t_r^{(s)}x^{m-2k}\,,
\ee
begin $[x]$ the integer part of $x$.
Using the standard relationships between eccentric anomaly $\psi$, true anomaly
$u$ and time $t$, see sec.~\ref{sec:orbital_motion}, one finds
\be
\ba{rcl}
1+e\cos\psi&=&\ds\frac{1-e^2}{1-e\cos u}\,,\\
d\beta&=&\ds\pa{1-e\cos u}du\,,
\ea
\ee
to obtain
\be
c_j^{(\ell,m)}=
\frac{2c_n}\pi\int_0^\pi
\sum_{k=0}^{[m/2]} t_m^{(k)}\frac{\pa{\cos u-e}^{m-2k}}{\pa{1-e\cos u}^{{m-2k}+\ell}}
\cos(ju-je\sin u)\,du\,.
\ee
In order to perform this integral we use the standard Taylor-expansions
\be
\ba{rcl}
(1-x)^n&=&\ds\sum_{k=0}^n (-1)^k\frac{n!}{k!(n-k)!}x^k\,,\\
\ds\frac 1{\pa{1-x}^n}&=&\ds\sum_{k=0}^\infty \frac{(n+k-1)!}{k!(n-1)!}x^k\,,
\ea
\ee
to write
\be
\resizebox{1.0 \textwidth}{!}{$
\ba{rcl}
c_j^{(\ell,m)}(e)&=&\ds
\frac{2c_j}{\pi}\int_0^\pi 
\sum_{k=0}^{[m/2]} t_m^{(k)}(-e)^{m-2k}\sum_{p=0}^{m-2k}
\frac{(m-2k)!}{p!(m-2k-p)!}(-1)^p\pa{\frac{\cos u}e}^p
\sum_{n=0}^\infty \frac{(m-2k+\ell+n-1)!}{n!(m-2k+\ell-1)!}\pa{e\cos u}^n\,
\cos(ju-je\sin u)\,du\,\\
&=&\ds \frac{2c_j}\pi\int_0^\pi \sum_{k=0}^{[m/2]}\sum_{p=0}^{m-2k}\sum_{n=0}^\infty
(-1)^{p+m}t_m^{(k)}
\frac{(m-2k)!}{p!(m-2k-p)!}\frac{(m-2k+\ell+n-1)!}{n!(m-2k+\ell-1)!}
e^{m-2k+n-p}\pa{\cos u}^{p+n}\cos(ju-je\sin u)\,du\,,
\ea $}
\ee
and then we use the DeMoivre formula
\be
\cos^n(u)=\frac 1{2^n}\sum_{k=0}^n\frac{n!}{k!(n-k)!}\cos(n-2k)u\,,
\ee
to get to
\be
\ba{rl}
\ds c_j^{(\ell,m)}(e)=\ds \frac{c_j}\pi\int_0^\pi
\sum_{k=0}^{[m/2]}\sum_{p=0}^{m-2k}\sum_{n=0}^\infty\sum_{q=0}^{p+n}
(-1)^{p+m}\frac{t_m^{(k)}}{2^{p+n-1}}&
\ds\frac{(m-2k)!}{p!(m-2k-p)!}\frac{(m-2k+\ell+n-1)!}{n!(m-2k+\ell-1)!}
\frac{(p+n)!}{q!(p+n-q)!}\\
&\ds \times e^{m-2k+n-p}\cos\paq{(p+n-2q)u}\cos(ju-je\sin u)\,du\,.
\ea
\ee
Finally using the integral representation of the Bessel functions
\be
J_n(z)=\frac 1\pi\int_0^\pi\cos(nu-z\sin u) du\,,
\ee
and the standard trigonometric identity
\be
2\cos\alpha \cos\beta=\cos(\alpha+\beta)+\cos(\alpha-\beta)\,,
\ee
one gets to
\be
\ba{rl}
c_j^{(\ell,m)}(e)=\ds c_j(-e)^m
\sum_{k=0}^{[m/2]}\sum_{p=0}^{m-2k}\sum_{n=0}^\infty\sum_{q=0}^{p+n}(-1)^p
e^{n-2k-p}\frac{t_m^{(k)}}{2^{p+n}}
\frac{(m-2k)!}{p!(m-2k-p)!}\frac{(m-2k+\ell+n-1)!}{n!(m-2k+\ell-1)!}
&\ds\frac{(p+n)!}{q!(p+n-q)!}\\
\ds\times\pa{J_{n+p+j-2q}(je)+J_{j-p-n+2q}(je)}\,.
\ea
\ee
Analogously for $s_j^{(\ell,m)}(e)$, one can use the Chebyshev polynomial of the
second kind $U_n(x)$ satisfying the equation
\be
\sin(m\theta)=U_{m-1}(\cos\theta)\sin\theta=
\sin\theta\sum_{k=0}^{[(m-1)/2]} u_{m-1}^{(k)}\pa{\cos\theta}^{m-1-2k}\,,
\ee
to obtain
\be
\ba{rcl}
s^{(\ell,m)}_j(e)&=&\ds \frac{2(1-e^2)^{1/2}}{\pi}\int_0^\pi\sin u
\sum_{k=0}^{[(m-1)/2]} u_{m-1}^{(k)}
\frac{\pa{\cos u-e}^{m-1-2k}}{\pa{1-e\cos u}^{m-2k+\ell}}
\sin(ju-je\sin u)\,du\\
&=&\ds-\frac{2(1-e^2)^{1/2}}\pi\int_0^\pi
\sum_{k=0}^{[(m-1)/2]}\sum_{p=0}^{m-1-2k}\sum_{n=0}^\infty\sum_{q=0}^{p+n}(-1)^{p+m}
e^{m-1-2k+n-p}\frac{u_m^{(k)}}{2^{p+n}}\\
&&\ds\times\frac{(m-2k-1)!}{p!(m-2k-1-p)!}\frac{(m-2k+\ell+n-1)!}{n!(m-2k+\ell-1)!}
\frac{(p+n)!}{q!(p+n-q)!}\sin u\cos\paq{(p+n-2q)u}\sin(ju-je\sin u)du\,.
\ea
\ee
Now using
\be
\ba{rcl}
2 \sin\alpha\cos\beta= \sin(\alpha+\beta)+\sin(\alpha-\beta)\,,\\
2 \sin\alpha\sin\beta=\cos(\alpha-\beta)-\cos(\alpha+\beta)\,,
\ea
\ee
one finally obtains
\be
\ba{rcl}
s_j^{(\ell,m)}(e)&=&\ds (1-e^2)^{1/2}(-e)^m
 \sum_{k=0}^{[(m-1)/2]}\sum_{p=0}^{m-1-2k}\sum_{n=0}^\infty\sum_{q=0}^{p+n}(-1)^p
e^{n-2k-p}\frac{u_{m-1}^{(k)}}{2^{p+n+1}}\\
&&\ds\times\frac{(m-2k-1)!}{p!(m-2k-1-p)!}\frac{(m-2k+\ell+n-1)!}{j!(m-2k+\ell-1)!}
\frac{(p+n)!}{q!(p+n-q)!}\\
&&\ds\times\paq{J_{n+p+j+1-2q}(je)+J_{j-p-n+1+2q}(je)-J_{n+p+j-1-2q}(je)-J_{j-p-n-1+2q}(je)}\,.
\ea
\ee

\section{Neutron star equations of state}
\label{sec:appC}
\noindent This appendix provides the numerical data for $f$-mode frequencies of four realistic equations of state.
In the first part of the  table of each equation of state we list the central density, the radius, the mass of the stellar model,  the frequencies of the $f$-mode for increasing values of $\ell$.
In the second part of each table we list the coefficients $|Q_{0\ell}|$.

\begin{table}
\begin{center}
\caption{Data for the equation of state A (APR) \protect\cite{Akmal:1998cf}, and \protect\cite{Haensel:2007wy} for the crust.}
\resizebox{0.70\textwidth}{!}{
\begin{tabular}{cccccccccccc}
\hline\hline
  $\rho_0$$ (\rm{gr}/\rm{cm}^3)$  & R(km)  & $M(M_\odot) $ & $\nu_f^{\ell=2}$(kHz) & $\nu_f^{\ell=3}$(kHz) & $\nu_f^{\ell=4}$(kHz)  & $|Q_{02}|$  &  $|Q_{03}|$  &   $|Q_{04}|$   & \\
\hline $1.5\times 10^{15}$ &  $11.132$   &  $1.965$  & $2.888$   &  $3.742$   &   $4.420$    & $2.321$  &  $2.437$  & $2.613$  &\\
\hline $1.2\times 10^{15}$ &  $11.433$   &  $1.704$  & $2.741$   &  $3.456$   &   $4.033$    & $2.258$  &  $2.482$  & $2.501$  &\\
\hline $9.9\times 10^{14}$ &  $11.603$   &  $1.408$  & $2.384$   &  $3.071$   &   $3.602$    & $2.323$  &  $2.594$  & $2.653$  &\\
\hline\hline
\label{table:1}
\end{tabular}}
\end{center}
\end{table}

\begin{table}
\begin{center}
\caption{ Data for the equation of state B (SLy4) \protect\cite{Douchin:2001sv} }
\resizebox{0.70\textwidth}{!}{
\begin{tabular}{cccccccccccc}
\hline\hline
  $\rho_0$$ (\rm{gr}/\rm{cm}^3)$  & R(km)  & $M(M_\odot) $ & $\nu_f^{\ell=2}$(kHz) & $\nu_f^{\ell=3}$(kHz) & $\nu_f^{\ell=4}$(kHz)  & $|Q_{02}|$  &  $|Q_{03}|$  &   $|Q_{04}|$    & \\
\hline $2.0\times 10^{15}$ &  $10.615$   &  $1.994$  &  $3.300$   &  $4.143$   &   $4.829$     &  $2.148$  &  $2.372$  &   $2.443$   &\\
\hline $1.6\times 10^{15}$ &  $11.017$   &  $1.884$  &  $3.024$   &  $3.808$   &   $4.461$     &  $2.180$  &  $2.439$  &   $2.446$   &\\
\hline $1.2\times 10^{15}$ &  $11.435$   &  $1.634$  &  $2.654$   &  $3.372$   &   $3.967$    &  $2.270$  &  $2.989$  &   $2.508$   &\\
\hline\hline
\label{table:2}
\end{tabular}}
\end{center}
\end{table}

\begin{table}
\begin{center}
\caption{ Data for the equation of state C \protect\cite{Walecka:1974qa} }
\resizebox{0.70\textwidth}{!}{
\begin{tabular}{cccccccccccc}
\hline\hline
  $\rho_0$$ (\rm{gr}/\rm{cm}^3)$  & R(km)  & $M(M_\odot) $ & $\nu_f^{\ell=2}$(kHz) & $\nu_f^{\ell=3}$(kHz) & $\nu_f^{\ell=4}$(kHz)  & $|Q_{02}|$ &  $|Q_{03}|$ &  $|Q_{04}|$   &\\
\hline $1.0\times 10^{15}$ &  $13.639$  &  $2.472$  &  $2.491$   &  $3.139$   &   $3.673$      & $2.093$   &  $2.365$     &  $2.465$      &\\
\hline $6.0\times 10^{14}$ &  $13.902$  &  $1.727$  &  $2.043$   &  $2.602$   &   $3.050$      & $2.281$   &  $2.300$     &  $2.449$      &\\
\hline $5.0\times 10^{14}$ &  $13.677$  &  $1.311$  &  $1.817$   &  $2.355$   &   $2.782$      & $2.261$   &  $2.351$     &  $2.622$      &\\
\hline\hline
\label{table:3}
\end{tabular}}
\end{center}
\end{table}

\begin{table}
\begin{center}
\caption{ Data for the equation of state D \protect\cite{Bethe:1974gy} }
\resizebox{0.70\textwidth}{!}{
\begin{tabular}{cccccccccccc}
\hline\hline
  $\rho_0$$ (\rm{gr}/\rm{cm}^3)$  & R(km)  & $M(M_\odot) $ & $\nu_f^{\ell=2}$(kHz) & $\nu_f^{\ell=3}$(kHz) & $\nu_f^{\ell=4}$(kHz)  & $|Q_{02}|$ &  $|Q_{03}|$ &  $|Q_{04}|$  & \\
\hline $1.6\times 10^{15}$ &  $11.131$  &  $1.691$  &  $2.842$   &  $3.593$   &   $4.202$    &   $2.130$     &  $2.609$     &  $2.576$     & \\
\hline $1.3\times 10^{15}$ &  $11.476$  &  $1.554$  &  $2.621$   &  $3.308$   &   $3.866$    &   $2.300$     &  $2.481$     &  $2.431$     & \\
\hline $1.2\times 10^{15}$ &  $11.620$  &  $1.488$  &  $2.517$   &  $3.151$   &   $3.660$    &   $2.249$    &  $2.426$     &  $2.406$     & \\
\hline\hline
\label{table:4}
\end{tabular}}
\end{center}
\end{table}

\section{Tensor spherical harmonics.}
\label{sec:appD}
The explicit expression of the tensor spherical harmonics
$\mathcal{Y}^{lm}_{i_1\ldots i_l}$ for $\ell=2$ are
\be
\mathcal{Y}^{22}_{ij}=\sqrt{\frac{15}{32\pi}}\left(
\ba{ccc}
1 & i & 0\\
i & -1 & 0\\
0 & 0 & 0
\ea
\right)_{ij}
\qquad
\mathcal{Y}^{21}_{ij}=-\sqrt{\frac{15}{32\pi}}\left(
\ba{ccc}
0 & 0 & 1\\
0 & 0 & i\\
1 & i & 0
\ea
\right)_{ij}
\qquad
\mathcal{Y}^{20}_{ij}=\sqrt{\frac{5}{16\pi}}\left(
\ba{ccc}
-1 & 0 & 0\\
0 & -1 & 0\\
0 & 0 & 2
\ea
\right)_{ij}
\ee
and $\mathcal{Y}^{2,-m}=(-1)^m{\mathcal{Y}^{2,m}}^*$.

\section{Energy and angular momentum absorption rates}

In this Appendix we report results for additional equations of state than the
one considered in the main text.
Figs.~\ref{fig:edotj_eos} show the distribution of energy absorbed $\dot E_j$
as a function of the fundamental mode frequency harmonic $j$ for the three
equations of state in tabs.~\ref{table:2},\ref{table:3},\ref{table:4}.

Figs.~\ref{fig:edote},\ref{fig:ldote}, display respectively the energy and
angular momentum absorbed by NS oscillations during the inspiral motion for
three different central density for each of the three equation of states reported in tabs.~\ref{table:2},\ref{table:3},\ref{table:4}.
For comparison the gravitational luminosity and angular momentum emitted in
gravitational wave are also reported.

\label{sec:appE}
\begin{figure}
    \includegraphics[width=.48\linewidth]{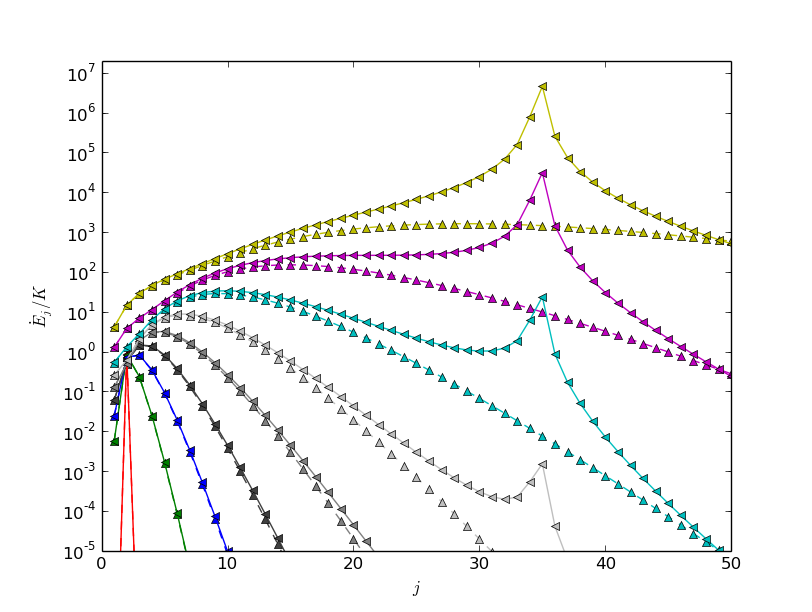}
    \includegraphics[width=.48\linewidth]{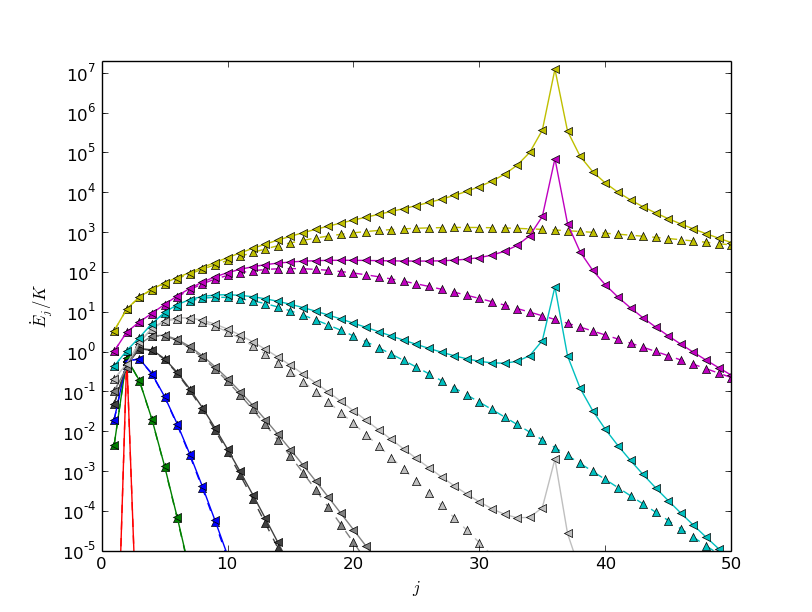}\\
    \includegraphics[width=.48\linewidth]{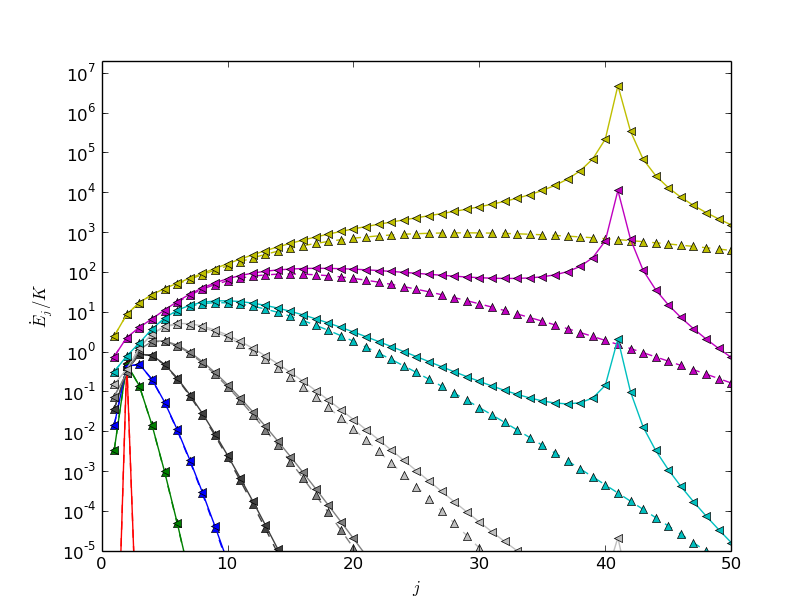}
\caption{Distribution of the energy per unit of mass absorbed by a single
NS oscillation $f-$mode $\dot E_j$ divided by the quantity $K$ defined in
eq.~(\ref{eq:Kdef}) as a function of the harmonic mode $j$ of the fundamental
orbital frequency in eccentric orbits. Going anti-clockwise from
top-left, the results are for the equation of state
\protect\cite{Douchin:2001sv} in tab.~\ref{table:2} for $\rho_0=2.0\cdot 10^{15}$ gr/cm$^{3}$,
\protect\cite{Walecka:1974qa} in tab.~\ref{table:3} for $\rho_0=1.0\cdot 10^{15}$ gr/cm$^{3}$,
\protect\cite{Bethe:1974gy} in tab.~\ref{table:4} for $\rho_0=1.6\cdot 10^{15}$ gr/cm$^{3}$.}
\label{fig:edotj_eos}
\end{figure}

\begin{figure}
  \includegraphics[width=.48\linewidth]{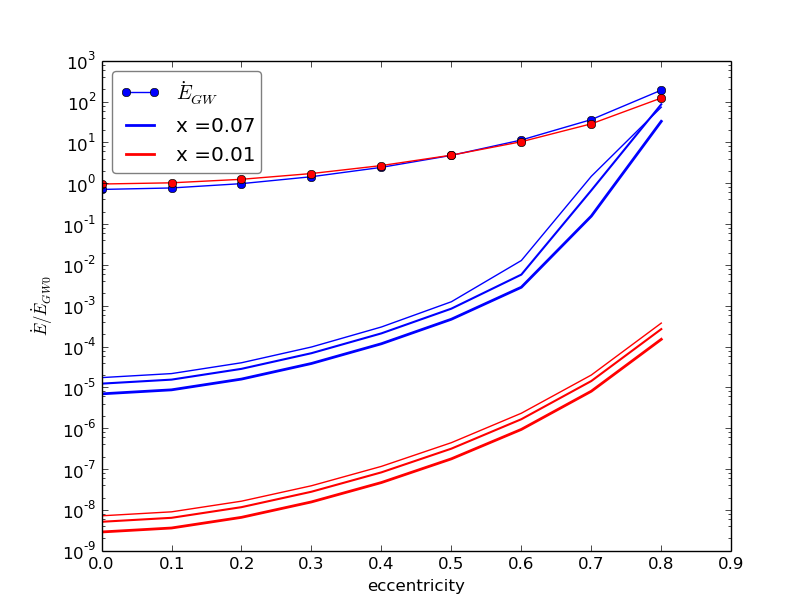}
  \includegraphics[width=.48\linewidth]{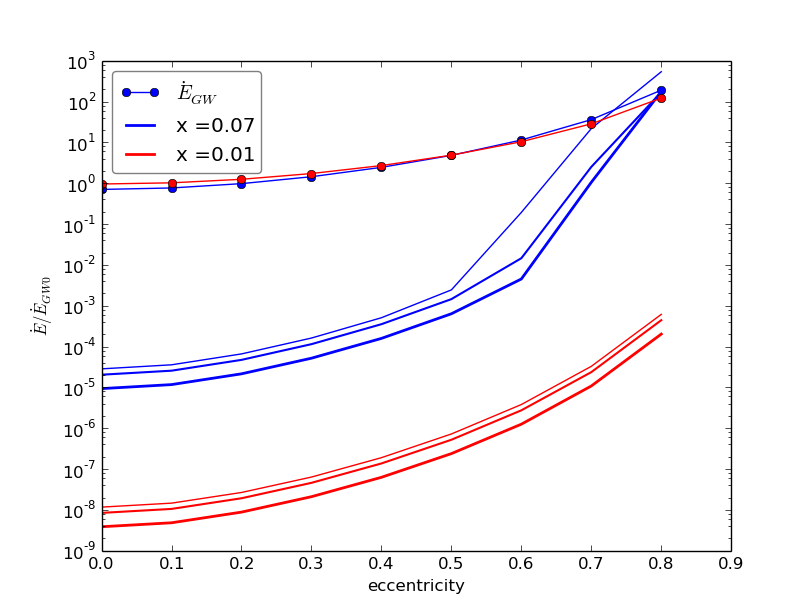}\\
  \includegraphics[width=.48\linewidth]{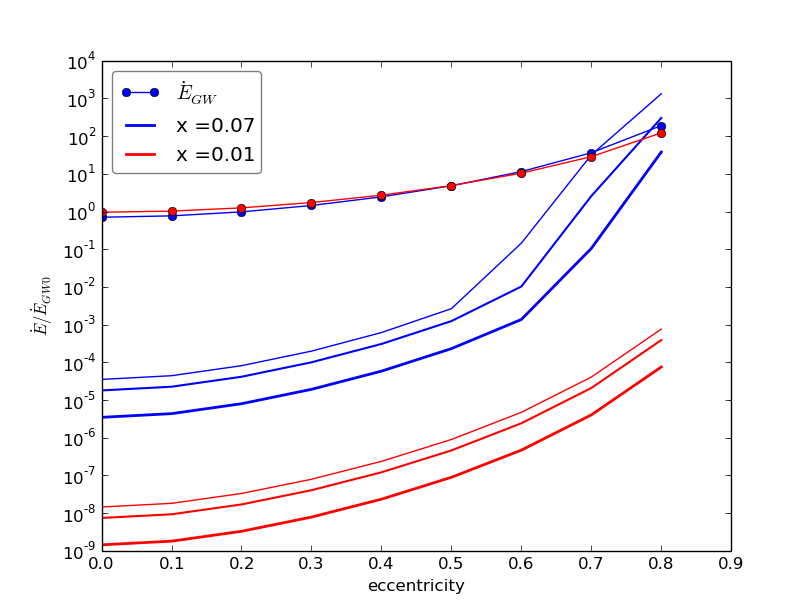}
\caption{Rate of energy absorbed $\dot E_*$ as a function of eccentricity, with
$M_{BH}=5M_\odot$, NS with equation of state respectively given, moving
anti-clockwise from top-left, by \protect\cite{Douchin:2001sv} in
tab.~\ref{table:2}, \protect\cite{Walecka:1974qa} in tab.~\ref{table:3},
and 
\protect\cite{Bethe:1974gy} in tab.~\ref{table:4}.
For comparison we also plot the GW luminosity for two values of $x$,
all functions are divided by the Newtonian GW luminosity at zero eccentricity
$\dot E_{GW0}$ given by eq.~(\ref{eq:Egw0}).
Note that for large eccentricity $e>0.7$ absorption by NS as
computed in this approximation is not negligible compared to GW emission.
For each equation of states results for the three values of the central density
reported in the corresponding tables are reported, increasing line thickness
denoting higher central density.}
\label{fig:edote_eos}
\end{figure}

\begin{figure}
\includegraphics[width=.48\linewidth]{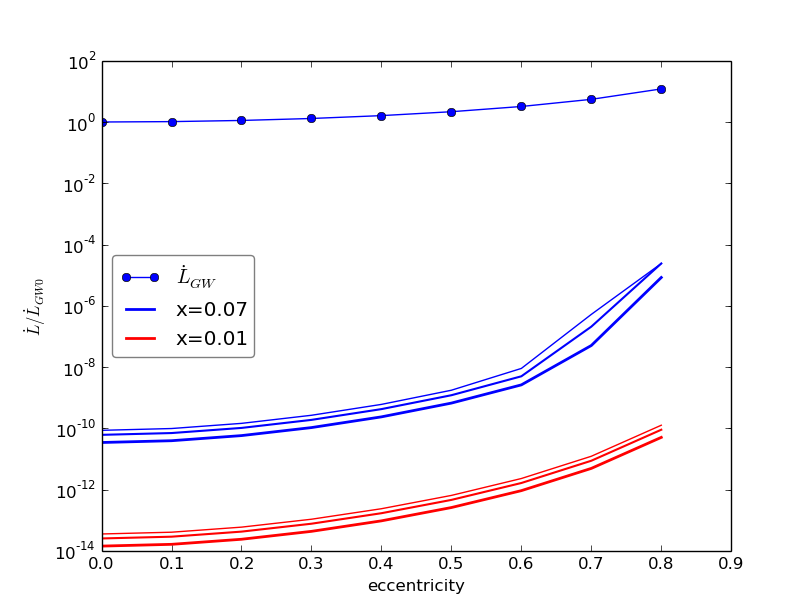}
\includegraphics[width=.48\linewidth]{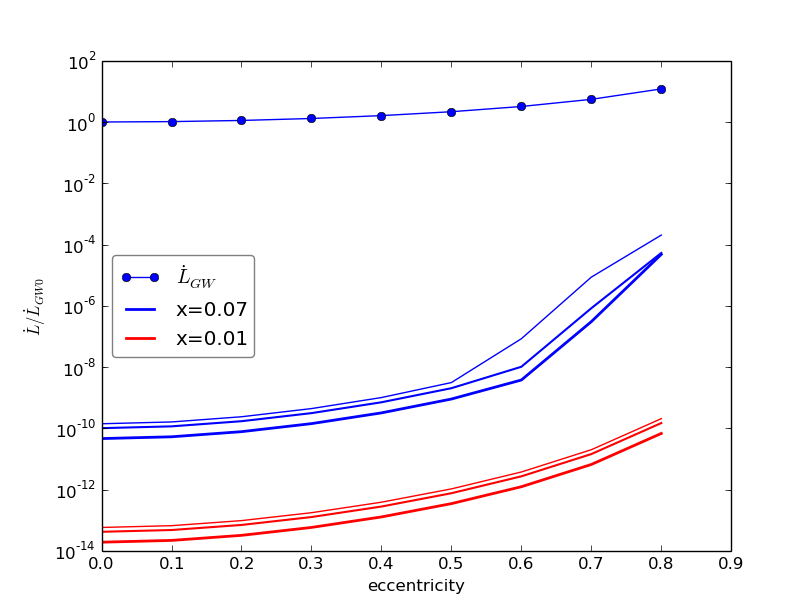}
\includegraphics[width=.48\linewidth]{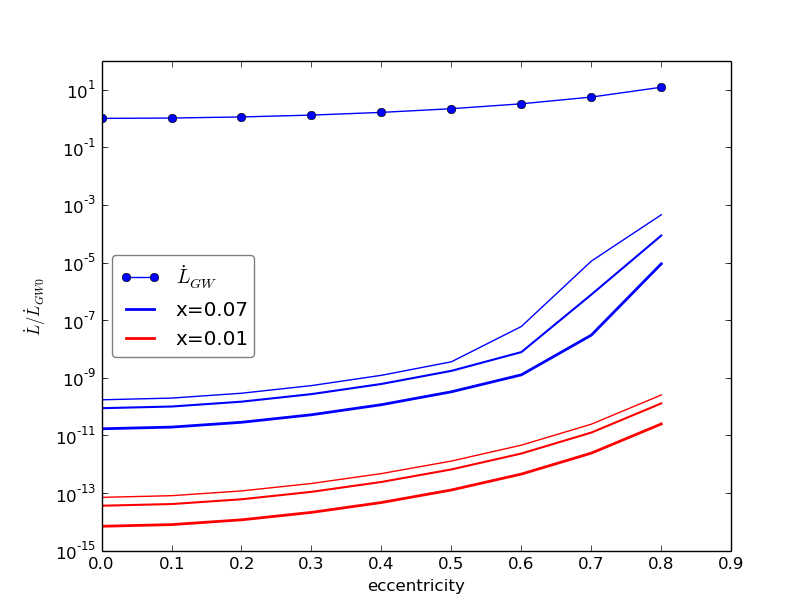}
\caption{Rate of angular momentum absorbed as a function of eccentricity, same
parameters as in fig.~\ref{fig:edote}. Here $\dot L_{GW}$ is the Newtonian angular
momentum loss in GWs for small eccentricities
$\dot L_{GW}=\frac{32}5\eta^2 M\frac{x^{7/2}}{(1-e^2)^2}\pa{1+\frac 78 e^2}$
and $\dot L_{GW0}=\dot L_{GW}|_{e=0}$, with $M\equiv M_*+M_{BH}$, $\eta\equiv M_*M_{BH}/M^2$.}
\label{fig:ldote_eos}
\end{figure}

\end{onecolumn}

\end{document}